\documentclass[preprintnumbers,amsmath,amssymb,floatfix,11pt,superscriptaddress,nofootinbib]{revtex4}
\usepackage{color}

\begin{document}
\title{Interacting entropy-corrected agegraphic-tachyon dark energy}

\author{ Mubasher Jamil} \email{mjamil@camp.nust.edu.pk}
\affiliation{Center for Advanced Mathematics and Physics, National
University of Sciences and Technology, H-12, Islamabad, Pakistan}

\author{Ahmad Sheykhi}
\email{sheykhi@mail.uk.ac.ir} \affiliation{Department of Physics,
Shahid Bahonar University, P.O. Box 76175, Kerman, Iran}

\begin{abstract}
\vspace*{1.5cm} \centerline{\bf Abstract} \vspace*{1cm} Motivated
by recent work of one of us \cite{ahmad}, we generalize this work
to agegraphic tachyon models of dark energy with entropy
correction terms arising from loop quantum gravity. We establish a
connection between the entropy-corrected agegraphic dark energy
and the tachyon scalar field in a universe with spacial curvature
and reconstruct the potential and the dynamics of the tachyon
scalar field which describe the tachyon cosmology. The
cosmological implications of the entropy-corrected agegraphic dark
energy models are also discussed.
\end{abstract}

\maketitle

\newpage
\section{Introduction}

A mysterious force propelling the universe, is one of the deepest
mysteries in all of science. This mysterious force now thought to
account for about seventy percent of the energy density of the
entire universe \cite{Sep} came to many as a surprise in 1998, when
the Supernova Cosmology Project and the High-z Supernova Search
teams \cite{Rie} independently announced their discovery that the
expansion of the universe is currently accelerating. Many theories
have been proposed to explain the cosmic acceleration expansion.
Although theories of trying to modify Einstein equations constitute
a big part of these attempts, the mainstream explanation for this
problem, however, is known as theories of dark energy. It is the
most accepted idea that a mysterious dominant component, dark
energy, with negative pressure, leads to this cosmic acceleration,
though the nature of such dark energy is still much source of doubt.

Many theoretical attempts towards understanding the dark energy
problem are focused to shed light on it in the framework of a
fundamental theory such as string theory or quantum gravity.
Although a complete theory of quantum gravity has not established
yet today, we still can make some attempts to investigate the
nature of dark energy according to some principles of quantum
gravity. The holographic dark energy (HDE) model and the
agegraphic dark energy (ADE)  model are just such examples, which
are originated from some considerations of the features of the
quantum theory of gravity. That is to say, the holographic and ADE
models possess some significant features of quantum gravity. The
former, that arose a lot of enthusiasm recently
\cite{Coh,Li,Huang,Hsu,jamil,HDE,Setare,Seta1,shey0}, is motivated
from the holographic hypothesis \cite{Suss1}. In Ref. \cite{wang},
a model of HDE with an interaction with matter fields had been
investigated by choosing the future event horizon as an IR cutoff.
It was shown that the ratio of energy densities can vary with
time. Moreover, with the interaction between the two different
constituents of the universe, they observed the evolution of the
universe, from early deceleration to late time acceleration. In
addition, they found that such an interacting dark energy model
could accommodate a transition of the dark energy from a normal
state where $w_D > -1$ to phantom regimes where $w_D < -1$.

The ADE is based on the uncertainty relation of quantum mechanics
together with the gravitational effect in general relativity. The
ADE model assumes that the observed dark energy comes from the
spacetime and matter field fluctuations in the universe
\cite{Cai1,Wei2,Wei1}. Following the line of quantum fluctuations of
spacetime, Karolyhazy \cite{kar} proposed that the distance $t$ in
Minkowski spacetime cannot be known to a better accuracy than
$\delta t=\lambda t_{p}^{2/3}t^{1/3}$, where $\lambda$ is a
dimensionless constant of order unity. Based on Karolyhazy relation,
Maziashvili proposed that the energy density of metric fluctuations
of Minkowski spacetime is given by \cite{maz}
\begin{equation}\label{2}
\rho_{D}\sim\frac{1}{t_{p}^{2}t^{2}}\sim\frac{M_{p}^{2}}{t^{2}},
 \end{equation}
 where $t_p$ and $M_{p}$ are the reduced Planck time and
 mass, respectively. Since in ADE model the
age of the universe is chosen as the length measure, instead of
the horizon distance, the causality problem in the HDE is avoided
\cite{Cai1}. The agegraphic models of dark energy  have been
examined and constrained by various astronomical observations
\cite{age,shey1,shey2,setare,age2}.

It is  important to note that the definition and derivation of
holographic energy density depends on the entropy-area relationship
of black holes in Einstein's gravity \cite{Coh}. However, this
definition can be modified from the inclusion of quantum effects,
motivated from the loop quantum gravity (LQG). The quantum
corrections provided to the entropy-area relationship leads to the
curvature correction in the Einstein-Hilbert action and vice versa
\cite{Zhu}. The corrected entropy takes the form \cite{modak}
\begin{equation}\label{S}
S=\frac{A}{4G}+\tilde\alpha \ln {\frac{A}{4G}}+\tilde\beta,
\end{equation}
where $\tilde\alpha$ and $\tilde\beta$ are dimensionless constants
of order unity. The exact values of these constants are not yet
determined and still is an open issue in loop quantum cosmology.
These corrections arise in the black hole entropy in LQG due to
thermal equilibrium fluctuations and quantum fluctuations
\cite{Rovelli}. Taking  the corrected entropy-area relation
(\ref{S}) into account, the energy density of the HDE will be
modified as well. On this basis, Wei \cite{wei} proposed
 the energy density of the so-called ``entropy-corrected holographic dark energy''
 (ECHDE) as
\begin{equation}\label{rhoS}
\rho _{D }=3n^2M_{p}^{2}L^{-2}+\alpha L^{-4}\ln
(M_{p}^{2}L^{2})+\beta L^{-4}.
\end{equation}
In this paper we would like to consider the so-called
``entropy-corrected agegraphic dark energy'' (ECADE) whose $L$ in
Eq. (\ref{rhoS}) is replaced with a time scale $t$ of the universe.
The energy density of ECADE is given by
\begin{equation}\label{rhoS1}
\rho_D=3n^2M_p^2t^{-2}+\alpha t^{-4}\ln(M_p^2t^2)+\beta t^{-4}.
\end{equation}
The motivation idea for taking the energy density of modified ADE
in the form (\ref{rhoS1}) comes from the fact that the origin of
ADE and HDE are the same. Indeed, it has been shown that the ADE
models are the HDE model with different IR length scales
\cite{Myung}. In the special case  $\alpha=\beta=0$, Eq.
(\ref{rhoS1}) yields the well-known agegraphic energy density
\cite{Cai1}. Since the last two terms in Eq. (\ref{rhoS1}) can be
comparable to the first term only when $t$ is very small, the
corrections make sense only at the early stage of the universe.
When the time scale $t$ becomes large, ECADE reduces to the
ordinary ADE.

On the other hand, among the various candidates to explain the
accelerated expansion, the rolling tachyon condensates in a class
of string theories may have interesting cosmological consequences.
The tachyon is an unstable field which has became important in
string theory through its role in the Dirac-Born-Infeld action
which is used to describe the D-brane action \cite{Sen1,Sen2}. It
has been shown  that the decay of D-branes produces a pressureless
gas with finite energy density that resembles classical dust
\cite{Sen3}.

Our aim here is to establish a correspondence between the
interacting ECADE scenarios and the tachyon scalar field in a
non-flat universe. Although it is generally believed that our
universe is flat, however, some experimental data has implied that
our universe is not a perfectly flat universe and recent papers
have favored the universe with spatial curvature \cite{spe}. We
suggest the entropy-corrected agegraphic description of the
tachyon dark energy and reconstruct the potential and the dynamics
of the tachyon scalar field which describe the tachyon cosmology.

This paper is organized as follows. In the next section we
associate the original ECADE with the tachyon field. In section
\ref{NEW}, we establish the correspondence between the new model
of interacting ECADE and the tachyon dark energy. In section
\ref{COS} we study the cosmological implications of ECADE models.
We summarize our results in section \ref{COS}.

\section{Tachyon reconstruction of original ECADE} \label{ORI}

The effective Lagrangian for the tachyon field is described by
\begin{eqnarray}
 L=-V(\phi)\sqrt{1-g^{\mu\nu}\partial_\mu \phi \partial_\nu \phi},
 \end{eqnarray}
where $V(\phi)$ is the tachyon potential. The corresponding energy
momentum tensor for the tachyon field can be written in a perfect
fluid form
\begin{eqnarray}
 T_{\mu\nu}=(\rho_\phi+p_\phi)u_{\mu} u_\nu-p_\phi g_{\mu\nu},
 \end{eqnarray}
where $\rho_\phi$ and $p_\phi$ are, respectively, the energy
density and pressure of the tachyon and the velocity $u_\mu$ is
\begin{eqnarray}
u_\mu=\frac{\partial_\mu \phi}{\sqrt{\partial_\nu \phi \partial^\nu
\phi}}.
 \end{eqnarray}
A rolling tachyon has an interesting equation of state whose
parameter smoothly interpolates between $-1$ and $0$ \cite{Gib1}.
Thus, tachyon can be realized as a suitable candidate for the
inflation at high energy \cite{Maz1} as well as a source of dark
energy depending on the form of the tachyon potential \cite{Padm}.
Therefore it becomes meaningful to reconstruct tachyon potential
$V(\phi)$ from some dark energy models possessing some significant
features of the quantum gravity theory, such as holographic and
ADE models. It was demonstrated that dark energy driven by
tachyon, decays to cold dark matter in the late accelerated
universe and this phenomenon yields a solution to cosmic
coincidence problem \cite{Sri}. The connection between tachyon
field and the HDE \cite{Setare4} and ADE \cite{ahmad,agetach} have
also been established.

We assume the background Friedmann-Robertson-Walker (FRW) universe
is described by the line element
\begin{equation}
ds^2=dt^2-a^2(t)\Big(\frac{dr^2}{1-kr^2}+r^2d\Omega^2\Big).
\end{equation}
where $a(t)$ is the scale factor, and $k$ is the curvature
parameter with $k = -1, 0, 1$ corresponding to open, flat, and
closed universes, respectively. A closed universe with a small
positive curvature ($\Omega_k\simeq0.01$) is compatible with
observations \cite{spe}. The first Friedmann equation takes the
form
\begin{eqnarray}\label{Fried}
H^2+\frac{k}{a^2}=\frac{1}{3M_p^2} \left( \rho_m+\rho_D \right),
\end{eqnarray}
where $\rho_m$ and $\rho_D$ are the energy densities of matter and
dark energy. The dimensionless density parameters are
\begin{equation}
\Omega_m=\frac{\rho_m}{\rho_{cr}},\ \ \
\Omega_D=\frac{\rho_D}{\rho_{cr}},\ \ \Omega_k=\frac{k}{a^2H^2},\
\ \rho_{cr}=\frac{3H^2}{8\pi G}.
\end{equation}
The last expression is the critical energy density. Using (10) in
(9), we get
\begin{equation}
1+\Omega_k=\Omega_m+\Omega_D.
\end{equation}
The energy density and pressure of the tachyon field are given by
\begin{equation}
\rho_\phi=-T_0^0=\frac{V(\phi)}{\sqrt{1-\dot\phi^2}},
\end{equation}
\begin{equation}
p_\phi=T_i^i=-V(\phi)\sqrt{1-\dot\phi^2}.
\end{equation}
The equation of state parameter is
\begin{equation}
w_\phi=\frac{p_\phi}{\rho_\phi}=\dot\phi^2-1.
\end{equation}
We assume the energy density of the original ECADE is of the form
(\ref{rhoS1}) where the time scale $t$ is chosen to be the age of
the universe,
\begin{equation}\label{rhoori}
\rho_D=3n^2M_p^2T^{-2}+\alpha T^{-4}\ln(M_p^2t^2)+\beta T^{-4},
\end{equation}
where $T$ is defined by
\begin{equation}
T=\int\limits_0^a\frac{da}{aH}.
\end{equation}
Using (10) and (15), we can write
\begin{equation}
\Omega_D=\frac{n^2}{H^2T^2}+\frac{\alpha}{3M_p^2H^2T^4}\ln(M_p^2T^2)
+\frac{\beta}{3M_p^2H^2T^4}.
\end{equation}
The energy conservation equation is
\begin{equation}
\dot\rho+3H(\rho+p)=0,\  \  \ \rho=\rho_m+\rho_D,\  \   p=p_D.
\end{equation}
Assuming an interaction of dark energy with dark  matter
\begin{eqnarray}
\dot\rho_m+3H\rho_m&=&Q,\\
\dot\rho_D+3H\rho_D(1+w_D)&=&-Q,
\end{eqnarray}
where $Q=3b^2H\rho$ is an energy exchange term and $b^2$ is a
coupling parameter \cite{jamil1}. Dark energy interacting with
dark matter is a promising model to solve the cosmic coincidence
problem. In Ref. \cite{wang1}, the authors studied the signature
of such interaction on large scale cosmic microwave background
(CMB) temperature anisotropies. Based on the detail analysis in
perturbation equations of dark energy and dark matter when they
are in interaction, they found that the large scale CMB,
especially the late Integrated Sachs Wolfe effect, is a useful
tool to measure the coupling between dark sectors. It was deduced
that in the 1$\sigma$ range, the constrained coupling between dark
sectors can solve the coincidence problem. In Ref. \cite{wang2}, a
general formalism to study the growth of dark matter perturbations
when dark energy perturbations and interactions between dark
sectors were presented. They showed that the dynamical stability
on the growth of structure depends on the form of coupling between
dark sectors. Moreover due to the influence of the interaction,
the growth index can differ from the value without interaction by
an amount up to the observational sensibility, which provides an
opportunity to probe the interaction between dark sectors through
future observations on the growth of structure.

Differentiating (15) with respect to cosmic time $t$, we obtain
\begin{equation}
\dot{\rho}_D=-2H\left(\frac{3n^2M_p^2T^{-2}+2\alpha
T^{-4}\ln(M_p^2T^2)+(2\beta-\alpha) T^{-4}}{\sqrt{3n^2M_p^2+\alpha
T^{-2}\ln(M_p^2T^2)+\beta T^{-2}}}\right)\sqrt{3M_p^2\Omega_D}.
\end{equation}
Using (21) in (20), we get
\begin{equation}
w_D=-1+\frac{2}{3}\left(\frac{3n^2M_p^2T^{-2}+2\alpha
T^{-4}\ln(M_p^2T^2)+(2\beta-\alpha) T^{-4}}{(3n^2M_p^2+\alpha
T^{-2}\ln(M_p^2T^2)+\beta
T^{-2})^{3/2}}\right)\sqrt{3M_p^2\Omega_D}-\frac{b^2(1+\Omega_k)}{\Omega_D}.
\end{equation}
Differentiating (17) with respect to cosmic time $t$ and using
$\dot\Omega_D=\Omega_D'H$, we obtain
\begin{eqnarray}
\Omega_D'&=&-2\frac{\dot
H}{H^4}\Big(\frac{n^2}{T^2}+\alpha\frac{\ln(M_p^2T^2)}{3M_p^2T^4}
+\beta\frac{1}{3M_p^2T^4} \Big)\nonumber\\&&-\frac{2}{(HT)^3}\Big(
n^2-\alpha\frac{1}{3M_p^2T^2}+\alpha\frac{2\ln(M_p^2T^2)}{3M_p^2T^2}+
\beta\frac{2}{3M_p^2T^2} \Big),
\end{eqnarray}
where the prime denotes differentiation w.r.t. $\ln a$, the
e-folding time parameter. Differentiation (9) w.r.t. $t$ and using
(15), (21) and (22), we get
\begin{eqnarray}
\frac{\dot H}{H^2}&=&\frac{-1}{3M_p^2H^2}\left(\frac{3n^2M_p^2
T^{-2}+2\alpha T^{-4}\ln(M_p^2 T^2)+(2\beta-\alpha)
T^{-4}}{\sqrt{3n^2M_p^2+\alpha T^{-2}\ln(M_p^2T^2)+\beta
T^{-2}}}\right)\sqrt{3M_p^2\Omega_D}\nonumber\\&&-\frac{\Omega_k}{2}-\frac{3}{2}(1-\Omega_D)
+\frac{3}{2}b^2(1+\Omega_k).
\end{eqnarray}
Using (17) and (24) in (23), we obtain
\begin{eqnarray}
\Omega_D'&=&\frac{-2}{H^2}\Big(\frac{n^2}{T^2}+\alpha\frac{\ln(M_p^2T^2)}{3M_p^2T^4}
+\beta\frac{1}{3M_p^2T^4} \Big)\Big[
\frac{-1}{3M_p^2H^2}\Big(\frac{3n^2M_p^2 T^{-2}+2\alpha
T^{-4}\ln(M_p^2 T^2)+(2\beta-\alpha) T^{-4}}{\sqrt{3n^2M_p^2+\alpha
T^{-2}\ln(M_p^2T^2)+\beta
T^{-2}}}\Big)\nonumber\\&&\times\sqrt{3M_p^2\Omega_D}-\frac{\Omega_k}{2}-\frac{3}{2}(1-\Omega_D)
+\frac{3}{2}b^2(1+\Omega_k) \Big]-2\Big(
 \frac{3M_p^2\Omega_D}{3n^2M_p^2+\alpha T^{-2}\ln(M_p^2 T^2)+\beta T^{-2}} \Big)^{3/2}\nonumber\\&&\times\Big(
n^2-\alpha\frac{1}{3M_p^2T^2}+\alpha\frac{2\ln(M_p^2T^2)}{3M_p^2T^2}+
\beta\frac{2}{3M_p^2T^2} \Big).
\end{eqnarray}
To develop correspondence between original ECADE and tachyon
field, we proceed
\begin{eqnarray}
V(\phi)&=&\rho_\phi\sqrt{1-\dot\phi^2},\\
&=&[3n^2M_p^2T^{-2}+\alpha T^{-4}\ln(M_p^2T^2)+\beta
T^{-4}]\nonumber\\&&\times\sqrt{1-\frac{2}{3}\Big(\frac{3n^2M_p^2T^{-2}+2\alpha
T^{-4}\ln(M_p^2T^2)+(2\beta-\alpha)T^{-4}}{(3n^2M_p^2+\alpha
T^{-2}\ln(M_p^2T^2)+\beta
T^{-2})^{3/2}}\Big)\sqrt{3M_p^2\Omega_D}+\frac{b^2(1+\Omega_k)}{\Omega_D}}.\nonumber\\
\end{eqnarray}
Also the kinetic energy gives
\begin{equation}
\dot\phi=\sqrt{1+w_D}=\sqrt{\frac{2}{3}\left(\frac{3n^2M_p^2T^{-2}+2\alpha
T^{-4}\ln(M_p^2T^2)+(2\beta-\alpha) T^{-4}}{(3n^2M_p^2+\alpha
T^{-2}\ln(M_p^2T^2)+\beta
T^{-2})^{3/2}}\right)\sqrt{3M_p^2\Omega_D}-\frac{b^2(1+\Omega_k)}{\Omega_D}}.
\end{equation}
Using $\dot\phi=\phi'H$, we can write
\begin{equation}
\phi'=\frac{1}{H}\sqrt{\frac{2}{3}\left(\frac{3n^2M_p^2T^{-2}+2\alpha
T^{-4}\ln(M_p^2T^2)+(2\beta-\alpha) T^{-4}}{(3n^2M_p^2+\alpha
T^{-2}\ln(M_p^2T^2)+\beta
T^{-2})^{3/2}}\right)\sqrt{3M_p^2\Omega_D}-\frac{b^2(1+\Omega_k)}{\Omega_D}}.
\end{equation}
Integration yields
\begin{equation}
\phi(a)-\phi(a_0)=\int\limits_{a_0}^a\frac{1}{aH}\sqrt{\frac{2}{3}\left(\frac{3n^2M_p^2T^{-2}+2\alpha
T^{-4}\ln(M_p^2T^2)+(2\beta-\alpha) T^{-4}}{(3n^2M_p^2+\alpha
T^{-2}\ln(M_p^2T^2)+\beta
T^{-2})^{3/2}}\right)\sqrt{3M_p^2\Omega_D}-\frac{b^2(1+\Omega_k)}{\Omega_D}}da.
\end{equation}
Alternatively we can write (30) as
\begin{equation}
\phi(t)-\phi(t_0)=\int\limits_{t_0}^t\sqrt{\frac{2}{3}\left(\frac{3n^2M_p^2T^{-2}+2\alpha
T^{-4}\ln(M_p^2T^2)+(2\beta-\alpha) T^{-4}}{(3n^2M_p^2+\alpha
T^{-2}\ln(M_p^2T^2)+\beta
T^{-2})^{3/2}}\right)\sqrt{3M_p^2\Omega_D}-\frac{b^2(1+\Omega_k)}{\Omega_D}}dt'.
\end{equation}
Here $a_0$  is the  value of the scale factor at the present time
$t_0$. Therefore, we have established an interacting
entropy-corrected agegraphic tachyon dark energy model and
reconstructed the potential and the dynamics of the tachyon field.
It is interesting to note that in the limiting case
$\alpha=0=\beta$, all the above expressions reduce to those
presented in \cite{ahmad}.

\section{Tachyon reconstruction of NEW ECADE}  \label{NEW}

The energy density of the new ECADE  is given by
\begin{equation}
\rho_D=3n^2M_p^2\eta^{-2}+\alpha \eta^{-4}\ln(M_p^2\eta^2)+\beta
\eta^{-4}.
\end{equation}
Here $\eta$ is the conformal time, which is defined by
\begin{equation}
\eta=\int\limits_0^a\frac{da}{a^2H}.
\end{equation}
Using (10) and (32), we can write
\begin{equation}
\Omega_D=\frac{n^2}{H^2\eta^2}+\frac{\alpha}{3M_p^2H^2\eta^4}\ln(M_p^2\eta^2)
+\frac{\beta}{3M_p^2H^2\eta^4}.
\end{equation}
Differentiating (32) with respect to cosmic time $t$, we obtain
\begin{equation}
\dot{\rho}_D=-\frac{2H}{a}\left(\frac{3n^2M_p^2\eta^{-2}+2\alpha
\eta^{-4}\ln(M_p^2\eta^2)+(2\beta-\alpha)
\eta^{-4}}{\sqrt{3n^2M_p^2+\alpha \eta^{-2}\ln(M_p^2\eta^2)+\beta
\eta^{-2}}}\right)\sqrt{3M_p^2\Omega_D}.
\end{equation}
Using (35) in (20), we get
\begin{equation}
w_D=-1+\frac{2}{3a}\left(\frac{3n^2M_p^2\eta^{-2}+2\alpha
\eta^{-4}\ln(M_p^2\eta^2)+(2\beta-\alpha)\eta^{-4}}{(3n^2M_p^2+\alpha
\eta^{-2}\ln(M_p^2\eta^2)+\beta
\eta^{-2})^{3/2}}\right)\sqrt{3M_p^2\Omega_D}-\frac{b^2(1+\Omega_k)}{\Omega_D}.
\end{equation}
Differentiating (34) with respect to cosmic time $t$ and using
$\dot\Omega_D=\Omega_D'H$, we obtain
\begin{eqnarray}
\Omega_D'&=&-2\frac{\dot
H}{H^4}\Big(\frac{n^2}{\eta^2}+\alpha\frac{\ln(M_p^2\eta^2)}{3M_p^2\eta^4}
+\beta\frac{1}{3M_p^2\eta^4}
\Big)\nonumber\\&&-\frac{2}{a(H\eta)^3}\Big(
n^2-\alpha\frac{1}{3M_p^2\eta^2}+\alpha\frac{2\ln(M_p^2\eta^2)}{3M_p^2\eta^2}+
\beta\frac{2}{3M_p^2\eta^2} \Big).
\end{eqnarray}
Differentiation (5) w.r.t. $t$ and using (16), (32) and (33), we get
\begin{eqnarray}
\frac{\dot
H}{H^2}&=&\frac{-1}{3aM_p^2H^2}\left(\frac{3n^2M_p^2\eta^{-2}+2\alpha
\eta^{-4}\ln(M_p^2\eta^2)+(2\beta-\alpha)
\eta^{-4}}{\sqrt{3n^2M_p^2+\alpha \eta^{-2}\ln(M_p^2\eta^2)+\beta
\eta^{-2}}}\right)\sqrt{3M_p^2\Omega_D}\nonumber\\&&-\frac{\Omega_k}{2}-\frac{3}{2}(1-\Omega_D)
+\frac{3}{2}b^2(1+\Omega_k).
\end{eqnarray}
Using (35) and (38) in (37), we obtain
\begin{eqnarray}
\Omega_D'&=&\frac{-2}{H^2}\Big(\frac{n^2}{\eta^2}+\alpha\frac{\ln(M_p^2\eta^2)}{3M_p^2\eta^4}
+\beta\frac{1}{3M_p^2\eta^4} \Big)\Big[
\frac{-1}{3aM_p^2H^2}\Big(\frac{3n^2M_p^2 \eta^{-2}+2\alpha
\eta^{-4}\ln(M_p^2 \eta^2)+(2\beta-\alpha)
\eta^{-4}}{\sqrt{3n^2M_p^2+\alpha \eta^{-2}\ln(M_p^2\eta^2)+\beta
\eta^{-2}}}\Big)\nonumber\\&&\times\sqrt{3M_p^2\Omega_D}-\frac{\Omega_k}{2}-\frac{3}{2}(1-\Omega_D)
+\frac{3}{2}b^2(1+\Omega_k) \Big]-\frac{2}{a}\Big(
 \frac{3M_p^2\Omega_D}{3n^2M_p^2+\alpha \eta^{-2}\ln(M_p^2 \eta^2)+\beta \eta^{-2}} \Big)^{3/2}\nonumber\\&&\times\Big(
n^2-\alpha\frac{1}{3M_p^2\eta^2}+\alpha\frac{2\ln(M_p^2\eta^2)}{3M_p^2\eta^2}+
\beta\frac{2}{3M_p^2\eta^2} \Big).
\end{eqnarray}

To develop correspondence between original ECADE and tachyon
field, we proceed
\begin{eqnarray}
V(\phi)&=&[3n^2M_p^2\eta^{-2}+\alpha \eta^{-4}\ln(M_p^2\eta^2)+\beta
\eta^{-4}]\nonumber\\&&\times\sqrt{1-\frac{2}{3a}\Big(\frac{3n^2M_p^2\eta^{-2}+2\alpha
\eta^{-4}\ln(M_p^2\eta^2)+(2\beta-\alpha)\eta^{-4}}{(3n^2M_p^2+\alpha
\eta^{-2}\ln(M_p^2\eta^2)+\beta
\eta^{-2})^{3/2}}\Big)\sqrt{3M_p^2\Omega_D}+\frac{b^2(1+\Omega_k)}{\Omega_D}}.\nonumber\\
\end{eqnarray}
Also the kinetic energy gives
\begin{equation}
\dot\phi=\sqrt{\frac{2}{3a}\left(\frac{3n^2M_p^2\eta^{-2}+2\alpha
\eta^{-4}\ln(M_p^2\eta^2)+(2\beta-\alpha)\eta^{-4}}{(3n^2M_p^2+\alpha
\eta^{-2}\ln(M_p^2\eta^2)+\beta
\eta^{-2})^{3/2}}\right)\sqrt{3M_p^2\Omega_D}-\frac{b^2(1+\Omega_k)}{\Omega_D}}.
\end{equation}
Using $\dot\phi=\phi'H$, we can write
\begin{equation}
\phi'=\frac{1}{H}\sqrt{\frac{2}{3a}\left(\frac{3n^2M_p^2\eta^{-2}+2\alpha
\eta^{-4}\ln(M_p^2\eta^2)+(2\beta-\alpha)\eta^{-4}}{(3n^2M_p^2+\alpha
\eta^{-2}\ln(M_p^2\eta^2)+\beta
\eta^{-2})^{3/2}}\right)\sqrt{3M_p^2\Omega_D}-\frac{b^2(1+\Omega_k)}{\Omega_D}}.
\end{equation}
Integration yields
\begin{equation}
\phi(a)-\phi(a_0)=\int\limits_{a_0}^a\frac{1}{aH}\sqrt{\frac{2}{3a}\left(\frac{3n^2M_p^2\eta^{-2}+2\alpha
\eta^{-4}\ln(M_p^2\eta^2)+(2\beta-\alpha)\eta^{-4}}{(3n^2M_p^2+\alpha
\eta^{-2}\ln(M_p^2\eta^2)+\beta
\eta^{-2})^{3/2}}\right)\sqrt{3M_p^2\Omega_D}-\frac{b^2(1+\Omega_k)}{\Omega_D}}da.
\end{equation}
Alternatively we can write (43) as
\begin{equation}
\phi(t)-\phi(t_0)=\int\limits_{t_0}^t\sqrt{\frac{2}{3a}\left(\frac{3n^2M_p^2\eta^{-2}+2\alpha
\eta^{-4}\ln(M_p^2\eta^2)+(2\beta-\alpha)\eta^{-4}}{(3n^2M_p^2+\alpha
\eta^{-2}\ln(M_p^2\eta^2)+\beta
\eta^{-2})^{3/2}}\right)\sqrt{3M_p^2\Omega_D}-\frac{b^2(1+\Omega_k)}{\Omega_D}}dt'.
\end{equation}
Again, one can easily check that in the special case
$\alpha=0=\beta$, all the above expressions reduce to those
discussed in \cite{ahmad}.

\section{Cosmological Implications of ECNADE models}  \label{COS}

As cosmological implications of the present model, we calculate
the effective equation of state parameter of interacting ECADE and
ECNADE. The effective equation of state parameter is defined by
\cite{jamil1}
\begin{equation}
w^{\text{eff}}_D=w_D+\frac{Q}{3H\rho_D}=w_D+b^2\Big(\frac{1+\Omega_k}{\Omega_D}\Big).
\end{equation}
Making use of Eqs. (22) and (36) in (45), we obtain
\begin{equation}
w_{\text{old}}^{\text{eff}}=-1+\frac{2}{3}\left(\frac{3n^2M_p^2T^{-2}+2\alpha
T^{-4}\ln(M_p^2T^2)+(2\beta-\alpha) T^{-4}}{(3n^2M_p^2+\alpha
T^{-2}\ln(M_p^2T^2)+\beta
T^{-2})^{3/2}}\right)\sqrt{3M_p^2\Omega_D},
\end{equation}
\begin{equation}
w_{\text{new}}^{\text{eff}}=-1+\frac{2}{3a}\left(\frac{3n^2M_p^2\eta^{-2}+2\alpha
\eta^{-4}\ln(M_p^2\eta^2)+(2\beta-\alpha)\eta^{-4}}{(3n^2M_p^2+\alpha
\eta^{-2}\ln(M_p^2\eta^2)+\beta
\eta^{-2})^{3/2}}\right)\sqrt{3M_p^2\Omega_D}.
\end{equation}
For the sake of completeness we also calculate the deceleration
parameter
\begin{equation}
q=-\frac{\ddot a}{aH^2}=-1-\frac{\dot H}{H^2}.
\end{equation}
Substituting (25) and (38) in (48) yields the expressions of
deceleration parameter for original and new ECADE, respectively:
\begin{eqnarray}
q_{\text{old}}&=&-1+\frac{1}{3M_p^2H^2}\left(\frac{3n^2M_p^2
T^{-2}+2\alpha T^{-4}\ln(M_p^2 T^2)+(2\beta-\alpha)
T^{-4}}{\sqrt{3n^2M_p^2+\alpha T^{-2}\ln(M_p^2T^2)+\beta
T^{-2}}}\right)\sqrt{3M_p^2\Omega_D}\nonumber\\&&+\frac{\Omega_k}{2}+\frac{3}{2}(1-\Omega_D)
-\frac{3}{2}b^2(1+\Omega_k), \\
q_{\text{new}}&=&-1+\frac{1}{3aM_p^2H^2}\left(\frac{3n^2M_p^2
\eta^{-2}+2\alpha \eta^{-4}\ln(M_p^2 \eta^2)+(2\beta-\alpha)
\eta^{-4}}{\sqrt{3n^2M_p^2+\alpha \eta^{-2}\ln(M_p^2\eta^2)+\beta
\eta^{-2}}}\right)\sqrt{3M_p^2\Omega_D}\nonumber\\&&+\frac{\Omega_k}{2}+\frac{3}{2}(1-\Omega_D)
-\frac{3}{2}b^2(1+\Omega_k).
\end{eqnarray}

\section{Summary and discussion}
In summary, among the various candidates to play the role of the
dark energy, tachyon has emerged as a possible source of dark
energy for a particular class of potentials \cite{Padm}. In this
paper, we have associated the interacting ECADE models with a
tachyon field which describe the tachyon cosmology in a non-flat
universe. The addition of correction terms to the energy density
of ADE models is a fundamental prediction of LQG and hence must be
taken into account while studying the dynamics of dark energy in
the universe. Using this modified energy density, we have
demonstrated that the universe can be described completely by a
tachyon scalar field in a certain way. We have adopted the
viewpoint that the scalar field models of dark energy are
effective theories of an underlying theory of dark energy. Thus,
we should be capable of using the scalar field model to mimic the
evolving behavior of the interacting ECADE and reconstructing this
scalar field model. We have reconstructed the potential and the
dynamics of the tachyon scalar field according to the evolutionary
behavior of the interacting ECADE models. We have also studied the
cosmological implications of ECADE scenarios. We would remark that
numerical values of the above effective EoSs and deceleration
parameters cannot be predicted from above due to ignorance of $T$
and $\eta$.

{
\subsubsection*{Acknowledgments}
We would like to thank the referees for giving very useful comments
to improve this work. }

\end{document}